\documentclass[12pt]{article}
\usepackage{epsfig}

\begin{document}

\begin{titlepage}

\begin{flushright}
CLNS~04/1903\\
{\tt hep-ph/0412241}\\[0.2cm]
December 16, 2004
\end{flushright}

\vspace{0.7cm}
\begin{center}
\Large\bf 
Two-Loop Relations for Heavy-Quark Parameters in the Shape-Function Scheme
\end{center}

\vspace{0.8cm}
\begin{center}
{\sc Matthias Neubert}\\
\vspace{0.7cm}
{\sl Institute for High-Energy Phenomenology\\
Newman Laboratory for Elementary-Particle Physics, Cornell University\\
Ithaca, NY 14853, U.S.A.\\[0.3cm]
and\\[0.3cm]
School of Natural Sciences, Institute for Advanced Study\\
Princeton, NJ 08540, U.S.A.}
\end{center}

\vspace{1.0cm}
\begin{abstract}
\vspace{0.2cm}\noindent
Moments of the renormalized $B$-meson shape function provide a natural way 
to define short-distance, running heavy-quark parameters such as the 
$b$-quark mass and kinetic energy. These parameters are particularly well 
suited for studies of inclusive decay distributions. The definitions of 
$m_b$ and $\mu_\pi^2$ in this ``shape-function scheme'' are derived to 
two-loop order. Using previous determinations of heavy-quark parameters in 
other schemes, we find $m_b(\mu_f)=(4.63\pm 0.08)$\,GeV and 
$\mu_\pi^2(\mu_f)=(0.15\pm 0.07)$\,GeV$^2$ at a reference scale 
$\mu_f=1.5$\,GeV.
\end{abstract}
\vfil

\end{titlepage}

\paragraph{\em Introduction.}
The past decade has seen a revolution in the precision with which fundamental 
physics can be probed using measurements of the decay properties of $b$ 
quarks. For instance, the element $|V_{cb}|$ of the quark mixing matrix is now 
known with a precision of 2\%, which is close to the accuracy of our knowledge 
of the Cabibbo angle \cite{Aubert:2004aw}. Measurements of rare decay 
processes such as $B\to X_s\gamma$ impose stringent constraints on model 
building. The $b$-quark mass has been determined with an uncertainty of about 
60\,MeV \cite{Aubert:2004aw}, which is much less than the QCD scale. To 
achieve such precision requires that heavy-quark parameters can be defined 
unambiguously using short-distance techniques. In particular, these 
definitions should not refer to the notion of on-shell quark states, which 
would introduce uncontrollable uncertainties (``renormalon ambiguities'') that 
far exceed the experimental precision achieved at the $B$ factories.

The question of a short-distance definition of the $b$-quark mass has received 
much attention. While an ad hoc subtraction scheme such as $\overline{\rm MS}$ 
in principle defines the quark mass in an unambiguous way, such a definition 
is not appropriate for the discussion of $B$ decays, where the typical scales 
are often significantly below $m_b$. A more fruitful concept is that of a 
low-scale subtracted quark mass \cite{Bigi:1996si}, which is based on the idea 
that non-perturbative contributions to the heavy-quark pole mass can be 
subtracted by making contact to some physical observable. The result is an 
expression $m_b(\mu_f)$, which differs from the pole mass by an amount 
proportional to a subtraction scale $\mu_f=\mbox{few}\times\Lambda_{\rm QCD}$. 

Several examples of such low-scale subtracted quark masses have been discussed 
in the literature. In the potential-subtraction scheme, long-distance 
contributions to the pole mass are subtracted by relating it to the static 
potential between two heavy quarks \cite{Beneke:1998rk}. Similarly, in the 
$\Upsilon(1S)$ scheme the quark mass is related to the mass of the 
lowest-lying bottomonium resonance \cite{Hoang:1998hm}. Both schemes are well 
suited to study heavy-quark systems in the non-relativistic regime, such as 
$b\bar b$ spectroscopy or heavy-quark production near threshold. A quark-mass 
definition more closely related to the non-perturbative physics probed in 
$B$-meson decays is provided by the kinetic scheme 
\cite{Bigi:1996si,Czarnecki:1997wy,Czarnecki:1997sz}, in which the 
non-perturbative subtraction is accomplished with the help of heavy-quark sum 
rules \cite{Bigi:1994ga}. Because these sum rules constrain the properties of 
$B$-meson form factors in the ``small-velocity limit'', the kinetic scheme is 
well suited for heavy-quark expansions applied to $B$ decays into charm 
particles. 

Inclusive $B$ decays into final states consisting of only light hadrons, such 
as $B\to X_s\gamma$ or $B\to X_u\,l\,\nu$, probe yet different aspects of 
non-perturbative physics. The bound-state effects relevant in these processes 
are encoded in $B$-meson shape functions defined in terms of the forward 
matrix elements of non-local string operators on the light cone 
\cite{Neubert:1993ch,Neubert:1993um,Bigi:1993ex}. A sensible definition of 
heavy-quark parameters for such processes should incorporate the bulk 
properties of the leading-order shape function $S(\omega)$. Schematically, 
inclusive decay spectra are given in terms of convolution integrals of the 
form 
\begin{equation}
   \int d\omega\,f(m_b^{\rm pole}+\omega)\,S(\omega)
   \approx f(m_b^{\rm pole}+\langle\omega\rangle)
   \equiv f(m_b^{\rm SF}) \,,
\end{equation}
where $f$ is some distribution, and in the last step we have defined the 
``shape-function mass'' by adding the average value of $\omega$ to the pole 
mass. More precisely, the basis of the shape-function scheme proposed in 
\cite{Bosch:2004th} is to obtain short-distance definitions of all heavy-quark 
parameters that enter the description of inclusive decay rates via the moments
\begin{equation}\label{MNdef}
   M_N(\mu_f,\mu) = \int_{-\mu_f}^{\infty}\!d\omega\,\omega^N\,S(\omega,\mu) 
\end{equation}
of the renormalized $B$-meson shape function $S(\omega,\mu)$, which has 
support over the range $-\infty<\omega\le m_B-m_b$. At tree level, the shape 
function is normalized to unity, its first moment vanishes (in the pole 
scheme), and higher moments are given in terms of local operator matrix 
elements in heavy-quark effective theory, e.g.\ $M_2^{\rm tree}=-\lambda_1/3$ 
\cite{Neubert:1993ch,Bigi:1993ex}, where $\lambda_1$ is the kinetic-energy 
parameter defined in the pole scheme \cite{Falk:1992wt}. While it is well 
known that the $b$-quark pole mass suffers from a renormalon ambiguity, the 
same is true for other heavy-quark parameters such as $\lambda_1$ 
\cite{Neubert:1996zy}, even though this may not be obvious from low-order
perturbative calculations \cite{Martinelli:1995zw}. Beyond the tree 
approximation, the moments $M_N(\mu_f,\mu)$ depend on the renormalization 
scale $\mu$ and the lower cutoff $\mu_f$ applied to the integral in 
(\ref{MNdef}), and it is natural to use them to define a set of running 
heavy-quark parameters order by order in perturbation theory. In fact, for 
$\mu_f\gg\Lambda_{\rm QCD}$ these moments can be calculated using an operator 
product expansion, in which the scale $\mu_f$ plays the role of a hard 
Wilsonian cutoff. The moment integrals can then be matched onto a series of 
forward $B$-meson matrix elements of local operators. The Wilson coefficients 
in this matching can be obtained using perturbation theory with free quark and 
gluon states. In particular, if the matching is performed in the parton model 
with on-shell heavy-quark states ($v\cdot k=0$, where $k=p_b-m_b\,v$ is the 
residual momentum, and $v$ denotes the $B$-meson velocity), then the local 
operator matrix elements are given by their tree-level expressions, and the 
matching becomes trivial once the moments (\ref{MNdef}) are computed to a 
given order in $\alpha_s$.

\paragraph{\em Two-loop relations in the shape-function scheme.}
The parton-model expression for the renormalized shape function was derived at 
one-loop order in \cite{Bosch:2004th,Bauer:2003pi}. The result involves 
so-called star distributions \cite{DeFazio:1999sv}, which are generalizations 
of plus distributions appropriate for test functions defined on an unbounded 
interval. This complication can be circumvented by introducing the function
\begin{equation}
   s(L,\mu) = \int_{-\mu_f}^{\infty}\!d\omega\,
   S_{\rm parton}(\omega,\mu) \,; \quad
   L = \ln\frac{\mu_f+n\cdot k}{\mu} \,,
\end{equation}
whose dependence on the cutoff $\mu_f$ enters only through the logarithm $L$, 
while $\mu$ dependence also resides in the coupling $\alpha_s(\mu)$. Here $n$ 
is a light-like vector ($n^2=0$, $v\cdot n=1$), which enters in the definition 
of the shape function as the matrix element of a non-local light-cone string 
operator \cite{Neubert:1993um}. The shape-function moments can now be obtained 
as
\begin{equation}\label{MNrela}
   M_N(\mu_f,\mu) = (-1)^N \left[ \mu_f^N\,s(L,\mu)
   - N \int_{-n\cdot k}^{\mu_f}\!d\sigma\,\sigma^{N-1}\,
   s\bigg( \ln\frac{\sigma+n\cdot k}{\mu},\mu\bigg) \right] .
\end{equation}
At the end of the calculation one must expand the result in powers of 
$n\cdot k$ and identify $n\cdot k\to 0$, $(n\cdot k)^2\to-\lambda_1/3$, etc.\ 
\cite{Bosch:2004th}.

The function $s(L,\mu)$ obeys an integro-differential renormalization-group 
equation, which can be derived starting from the evolution equation for the
shape function obtained in \cite{Bosch:2004th,Neubert:2004dd}. We find
\begin{eqnarray}\label{RGE}
   \frac{d}{d\ln\mu}\,s(L,\mu)
   &=& 2 \Big[ \Gamma_{\rm cusp}(\alpha_s)\,L - \gamma(\alpha_s) \Big]\,
    s(L,\mu) \nonumber\\
   &&\mbox{}+ 2 \Gamma_{\rm cusp}(\alpha_s)
    \int_0^1 \frac{dz}{z} \Big[ s(L+\ln(1-z),\mu) - s(L,\mu) \Big] \,,
\end{eqnarray}
where $\alpha_s\equiv\alpha_s(\mu)$, and $\Gamma_{\rm cusp}$, $\gamma$ are
anomalous dimensions. This equation can be solved order by order in 
perturbation theory with an ansatz of the form
\begin{equation}
   s(L,\mu) = 1 + \sum_{n=1}^\infty
   \left( \frac{\alpha_s(\mu)}{4\pi} \right)^n
   \left( c_0^{(n)} + c_1^{(n)} L + \dots + c_{2n-1}^{(n)} L^{2n-1}
   + c_{2n}^{(n)} L^{2n} \right) .
\end{equation}
The evolution equation (\ref{RGE}) allows us to express all coefficients of 
logarithms in terms of the perturbative expansion coefficients of the 
anomalous dimensions and $\beta$ function, defined as
\begin{eqnarray}
   \Gamma_{\rm cusp}(\alpha_s)
   &=& \sum_{n=0}^\infty \Gamma_n
    \left( \frac{\alpha_s}{4\pi} \right)^{n+1} , \qquad
   \gamma(\alpha_s) = \sum_{n=0}^\infty \gamma_n
    \left( \frac{\alpha_s}{4\pi} \right)^{n+1} , \nonumber\\
   \beta(\alpha_s) &=& \frac{d\alpha_s}{d\ln\mu}
    = -2\alpha_s \sum_{n=0}^\infty \beta_n
    \left( \frac{\alpha_s}{4\pi} \right)^{n+1} .
\end{eqnarray}
At two-loop order, we obtain
\begin{eqnarray}\label{ansatz}
   s(L,\mu) &=& 1 + \frac{\alpha_s(\mu)}{4\pi} \left[
    c_0^{(1)} + 2\gamma_0\,L - \Gamma_0\,L^2 \right] \nonumber\\
   &&\mbox{}+ \left( \frac{\alpha_s(\mu)}{4\pi} \right)^2 \Bigg[
    c_0^{(2)} + \left( 2c_0^{(1)} (\gamma_0-\beta_0) + 2\gamma_1
    + \frac{2\pi^2}{3}\,\Gamma_0\gamma_0 + 4\zeta_3\,\Gamma_0^2 \right) L \\
   &&\mbox{}+ \left( 2\gamma_0(\gamma_0-\beta_0) - c_0^{(1)}\Gamma_0
    - \Gamma_1 - \frac{\pi^2}{3}\,\Gamma_0^2 \right) L^2 
    + \left( \frac23\,\beta_0 - 2\gamma_0 \right) \Gamma_0\,L^3
    + \frac12\,\Gamma_0^2\,L^4 \Bigg] \,. \nonumber
\end{eqnarray}
The relevant expansion coefficients of the $\beta$ function and cusp anomalous 
dimension $\Gamma_{\rm cusp}$ \cite{Korchemsky:1987wg} are (in the 
$\overline{\rm MS}$ renormalization scheme)
\begin{equation}
   \beta_0 = \frac{11}{3}\,C_A - \frac23\,n_f \,, \qquad
   \Gamma_0 = 4 C_F \,, \qquad
   \Gamma_1 = C_F \left[ \left( \frac{268}{9} - \frac{4\pi^2}{3} \right)
    C_A - \frac{40}{9}\,n_f \right] .
\end{equation}
The two-loop coefficient of the anomalous dimension $\gamma$ has been 
calculated in \cite{Korchemsky:1992xv}. We have found some mistakes in the 
translation of the results for the two-loop graphs into the expression for the 
anomalous dimension. The corrected result is \cite{Neubert:2004dd,privcom}
\begin{equation}\label{littlegamma}
   \gamma_0 = - 2 C_F \,, \qquad
   \gamma_1 = C_F \left[ \left( \frac{74}{27} + \frac{\pi^2}{18}
    - 18\zeta_3 + \kappa \right) C_A
    + \left( \frac{4}{27} + \frac{\pi^2}{9} \right) n_f \right] \,,
\end{equation}
where $\kappa=0$ under the assumption that the two-loop diagrams themselves 
were evaluated correctly in \cite{Korchemsky:1992xv}. However, there is reason
to believe that there might be an additional error in that paper, giving rise 
to a non-zero value $\kappa=4/3$ \cite{Einanprivate}, which we adopt in our 
numerical analysis. The expression for $\gamma_1$ should be checked with an 
independent calculation. Finally, the coefficient $c_0^{(1)}=(-\pi^2/6)\,C_F$ 
was computed in \cite{Bosch:2004th,Bauer:2003pi}, while the coefficient 
$c_0^{(2)}$ is presently unknown. This coefficient does not enter the 
definitions of heavy-quark parameters at two-loop order in the shape-function 
scheme.

Using the explicit two-loop expression (\ref{ansatz}) for the function 
$s(L,\mu)$, it is straightforward to compute expressions for the 
shape-function moments (\ref{MNrela}) in the pole scheme. Closed expressions
valid at one-loop order are given in \cite{Bosch:2004th}. One finds that the 
first moment $M_1(\mu_f,\mu)$ no longer vanishes beyond the tree approximation 
but receives a perturbative correction proportional to $\mu_f\,\alpha_s(\mu)$. 
The key idea behind the shape-function scheme is to redefine the $b$-quark 
mass, $m_b^{\rm pole}\equiv m_b(\mu_f,\mu)+\delta m$, in such a way that after 
the redefinition the first moment vanishes. Technically, this is implemented 
by introducing a residual mass term $\delta m$ for the heavy quark 
\cite{Falk:1992fm}. All that changes in the expressions for the moments is 
that the residual momentum $k$ gets replaced by $k+\delta m\,v$, and hence 
$n\cdot k\to n\cdot k+\delta m$. We then choose $\delta m$ such that 
$M_1(\mu_f,\mu)=0$, thereby defining a mass renormalization scheme order by 
order in perturbation theory. Next, we define the kinetic-energy parameter 
$\mu_\pi^2(\mu_f,\mu)$ via the ratio $3M_2(\mu_f,\mu)/M_0(\mu_f,\mu)$ in the 
new scheme. This quantity differs from the pole-scheme parameter $-\lambda_1$ 
by terms of order $\mu_f^2\,\alpha_s(\mu)$. We then solve for $m_b^{\rm pole}$ 
and $\lambda_1$ in terms of the new parameters $m_b(\mu_f,\mu)$ and 
$\mu_\pi^2(\mu_f,\mu)$. To two-loop order, the results of these manipulations 
are
\begin{eqnarray}\label{beauty}
   m_b^{\rm pole} 
   &=& m_b(\mu_f,\mu) + \mu_f\,\frac{C_F\alpha_s(\mu)}{\pi}
    \left[ 1 - 2\ln\frac{\mu_f}{\mu} + \frac{\alpha_s(\mu)}{\pi}\,
    k_1(\mu_f,\mu) \right] \nonumber\\
   &&\mbox{}+ \frac{\mu_\pi^2(\mu_f,\mu)}{3\mu_f}\,
    \frac{C_F\alpha_s(\mu)}{\pi} \left[ 2\ln\frac{\mu_f}{\mu}
    + \frac{\alpha_s(\mu)}{\pi}\,k_2(\mu_f,\mu) \right] , \nonumber\\
   - \lambda_1
   &=& \mu_\pi^2(\mu_f,\mu) \left[ 1 + \frac{C_F\alpha_s(\mu)}{\pi}
    \left( - \frac12 - 3\ln\frac{\mu_f}{\mu}
    + \frac{\alpha_s(\mu)}{\pi}\,k_3(\mu_f,\mu) \right) \right] \nonumber\\
   &&\mbox{}+ \mu_f^2\,\frac{C_F\alpha_s(\mu)}{\pi}
    \left[ 3\ln\frac{\mu_f}{\mu}
    + \frac{\alpha_s(\mu)}{\pi}\,k_4(\mu_f,\mu) \right] ,
\end{eqnarray}
where the one-loop terms agree with \cite{Bosch:2004th}, while the two-loop
corrections are encoded in the coefficient functions
\begin{eqnarray}\label{kires}
   k_1(\mu_f,\mu)
   &=& \frac{47}{36}\,\beta_0
    + \left( \frac{10}{9} - \frac{\pi^2}{12} - \frac94\,\zeta_3
    + \frac{\kappa}{8} \right) C_A
    + \left( - 8 + \frac{\pi^2}{3} + 4\zeta_3 \right) C_F \nonumber\\
   &&\mbox{}+ \left[ - \frac43\,\beta_0
    + \left( - \frac23 + \frac{\pi^2}{6} \right) C_A
    + \left( 8 - \frac{2\pi^2}{3} \right) C_F \right] \ln\frac{\mu_f}{\mu}
    + \left( \frac12\,\beta_0 + 2C_F \right) \ln^2\frac{\mu_f}{\mu} \,,
    \nonumber\\
   k_2(\mu_f,\mu)
   &=& - k_1(\mu_f,\mu)
    + \frac76\,\beta_0 + \left( \frac13 - \frac{\pi^2}{12} \right) C_A
    + \left( - 5 + \frac{\pi^2}{3} \right) C_F
    + \left( - \frac12\,\beta_0 - C_F \right) \ln\frac{\mu_f}{\mu} \,,
    \nonumber\\
   k_3(\mu_f,\mu)
   &=& \frac18\,\beta_0
    + \left( 1 + \frac{\pi^2}{24} - \frac{27}{8}\,\zeta_3
    + \frac{3\kappa}{16} \right) C_A
    + \left( - \frac{13}{4} - \frac{\pi^2}{6} + 6\zeta_3 \right) C_F
    \nonumber\\
   &&\mbox{}+ \left[ - \beta_0
    + \left( - 1 + \frac{\pi^2}{4} \right) C_A
    + \left( \frac{27}{2} - \pi^2 \right) C_F \right] \ln\frac{\mu_f}{\mu}
    + \left( \frac34\,\beta_0 + 2 C_F \right) \ln^2\frac{\mu_f}{\mu} \,,
    \nonumber\\
   k_4(\mu_f,\mu)
   &=& - k_3(\mu_f,\mu)
    - \frac{11}{24}\,\beta_0
    + \left( - \frac16 + \frac{\pi^2}{24} \right) C_A
    + \left( \frac72 - \frac{\pi^2}{6} \right) C_F \nonumber\\
   &&\mbox{}+ \left( \frac14\,\beta_0 - \frac32\,C_F \right)
    \ln\frac{\mu_f}{\mu} - 4 C_F \ln^2\frac{\mu_f}{\mu} \,.
\end{eqnarray}
Equations (\ref{beauty}) and (\ref{kires}) are the main results of this work. 
They allow us to consistently implement the shape-function scheme with 
two-loop accuracy in perturbative calculations of heavy-quark processes. They 
also enable us to connect the shape-function scheme with any other 
short-distance scheme used to define heavy-quark parameters.

For practical applications, it is often inconvenient to use heavy-quark 
parameters defined as a function of two scales, and we will therefore adopt 
the special choice $\mu=\mu_f$ as our default, denoting 
$m_b(\mu_f)\equiv m_b(\mu_f,\mu_f)$ and 
$\mu_\pi^2(\mu_f)\equiv\mu_\pi^2(\mu_f,\mu_f)$. Then the relations 
(\ref{beauty}) simplify, since all logarithms vanish. For $N_c=3$, we obtain 
\begin{eqnarray}\label{simplebeauty}
   m_b^{\rm pole} 
   &=& m_b(\mu_f) + \mu_f\,\frac{4\alpha_s(\mu_f)}{3\pi}
    \left[ 1 + \frac{\alpha_s(\mu_f)}{\pi}\,
    \left( \frac{253}{36} + \frac{7\pi^2}{36} - \frac{17}{12}\,\zeta_3
    + \frac{3\kappa}{8} - \frac{47}{54}\,n_f \right) \right] \nonumber\\
   &&\mbox{}+ \frac{\mu_\pi^2(\mu_f)}{\mu_f}
    \left( \frac{\alpha_s(\mu_f)}{\pi} \right)^2
    \left( \frac{5}{81} + \frac{17}{27}\,\zeta_3 - \frac{\kappa}{6}
    + \frac{10}{243}\,n_f \right) , \nonumber\\
   - \lambda_1
   &=& \mu_\pi^2(\mu_f) \left[ 1 - \frac{2\alpha_s(\mu_f)}{3\pi}
    + \left( \frac{\alpha_s(\mu_f)}{\pi} \right)^2 
    \left( \frac{1}{18} - \frac{7\pi^2}{54} - \frac{17}{6}\,\zeta_3
    + \frac{3\kappa}{4} - \frac{n_f}{9} \right) \right] \nonumber\\
   &&\mbox{}+ \mu_f^2 \left( \frac{\alpha_s(\mu_f)}{\pi} \right)^2
    \left( - \frac{11}{9} + \frac{17}{6}\,\zeta_3 - \frac{3\kappa}{4}
    + \frac{14}{27}\,n_f \right) .
\end{eqnarray}

\paragraph{\em Connections with other schemes.}
To the best of our knowledge, the only other scheme in which both $m_b$ and 
$\mu_\pi^2$ are defined in a consistent way (without renormalon ambiguities) 
is the kinetic scheme \cite{Bigi:1996si}. The relevant two-loop relations in 
that scheme are \cite{Czarnecki:1997wy,Czarnecki:1997sz}
\begin{eqnarray}\label{kinscheme}
   m_b^{\rm pole}
   &=& m_b(\mu_f) + \frac43\,\mu_f\,\frac{C_F\alpha_s(2\mu_f)}{\pi}
    \left\{ 1 + \frac{\alpha_s(2\mu_f)}{\pi} \left[ \frac43\,\beta_0
    + \left( \frac{13}{12} - \frac{\pi^2}{6} \right) C_A \right] \right\}
    \nonumber\\
   &&\mbox{}+ \frac{\mu_f^2}{2m_b(\mu_f)}\,\frac{C_F\alpha_s(2\mu_f)}{\pi}
    \left\{ 1 + \frac{\alpha_s(2\mu_f)}{\pi} \left[ \frac{13}{12}\,\beta_0
    + \left( \frac{13}{12} - \frac{\pi^2}{6} \right) C_A \right] \right\} ,
    \nonumber\\
   - \lambda_1
   &=& \mu_\pi^2(\mu_f) - \mu_f^2\,\frac{C_F\alpha_s(2\mu_f)}{\pi}
    \left\{ 1 + \frac{\alpha_s(2\mu_f)}{\pi} \left[ \frac{13}{12}\,\beta_0
    + \left( \frac{13}{12} - \frac{\pi^2}{6} \right) C_A \right] \right\} . 
\end{eqnarray}
The fact that the $b$-quark mass receives a $1/m_b$ correction in the 
kinetic scheme leads to a mixing between different terms in the $1/m_b$ 
expansion. No such mixing occurs in the shape-function scheme. Other 
approaches such as the potential-subtraction or $\Upsilon(1S)$ schemes only 
redefine the $b$-quark mass but ignore the problem of infrared ambiguities of 
other heavy-quark parameters. The two-loop relation between the 
potential-subtracted quark mass and the pole mass reads \cite{Beneke:1998rk}
\begin{equation}\label{PSmassdef}
   m_b^{\rm pole} = m_b(\mu_f) + \mu_f\,\frac{C_F\alpha_s(\mu_f)}{\pi}
    \left[ 1 + \frac{\alpha_s(\mu_f)}{\pi}
    \left( \frac{11}{12}\,\beta_0 - \frac23\,C_A \right) \right] .
\end{equation}
At one-loop order, the potential-subtracted quark mass coincides with the mass 
defined (at the same scale $\mu_f$) in the shape-function scheme. The 
next-to-leading order relation between the pole mass and the $b$-quark mass 
defined in the $\Upsilon(1S)$ scheme is \cite{Hoang:1999zc}
\begin{equation}\label{1Smassdef}
   m_b = m_b^{\rm pole} - \frac{[C_F\alpha_s(\mu)]^2}{8}\,m_b^{\rm pole}
   \left\{ 1 + \frac{\alpha_s(\mu)}{\pi} \left[
   \left( L + \frac{11}{6} \right) \beta_0 - \frac43\,C_A \right] \right\} ,
\end{equation}
where $L=\ln[\mu/C_F\alpha_s(\mu)\,m_b^{\rm pole}]$. The peculiar counting of
powers of $\alpha_s(\mu)$ in the $\Upsilon(1S)$ scheme becomes more 
transparent if we rewrite this relation in the form
\begin{equation}
   m_b^{\rm pole} = m_b + \mu_f(\mu)\,\frac{C_F\alpha_s(\mu)}{\pi}
   \left\{ 1 + \frac{\alpha_s(\mu)}{\pi} \left[
   \left( \ln\frac{\pi\mu}{8\mu_f(\mu)} + \frac{11}{6} \right) \beta_0
   - \frac43\,C_A \right] \right\} , 
\end{equation}
where
\begin{equation}\label{funny}
   \mu_f(\mu) = \frac{\pi}{8}\,C_F\alpha_s(\mu)\,m_b^{\rm pole}
\end{equation}
plays the role of the subtraction scale. Note that the $\Upsilon(1S)$ mass is
scale independent, contrary to the other low-scale subtracted quark masses 
considered above. On the other hand, the subtraction point $\mu_f(\mu)$ is
itself a function of the renormalization scale $\mu$. Based on 
(\ref{1Smassdef}) one would suspect that a proper choice of $\mu$ is such
that $L\approx 0$, which yields $\mu\approx 2$\,GeV. However, relation 
(\ref{funny}) suggests that the appropriate physical scale in the 
$\Upsilon(1S)$ scheme may be lower, in accordance with arguments presented in 
\cite{Beneke:1999fe}.

\paragraph{\em Numerical results.}
Estimates of the $b$-quark mass and kinetic energy in the shape-function 
scheme can be obtained using various sources of phenomenological information 
derived in other subtraction schemes. For this purpose we adopt the scale 
choice $\mu_f=\mu=1.5$\,GeV in the shape-function scheme. The running of the 
heavy-quark parameters will be studied later. As mentioned above, we use 
$\kappa=4/3$.

The values of $m_b$ and $\mu_\pi^2$ defined in the kinetic scheme at a 
reference scale $\mu_f=1$\,GeV have recently be determined from a global fit 
to experimental data on moments of various $B\to X_c\,l\,\nu$ decay 
distributions \cite{Aubert:2004aw}, yielding $m_b=(4.611\pm 0.068)$\,GeV and 
$\mu_\pi^2=(0.447\pm 0.053)$\,GeV$^2$ (with a mild anti-correlation of errors, 
$c=-0.4$), in good agreement with theoretical expectations 
\cite{Gambino:2004qm}. Using these results, we compute the pole scheme 
parameters from (\ref{kinscheme}) and then solve for the parameters in the 
shape-function scheme using (\ref{simplebeauty}). We obtain
\begin{equation}\label{values}
   m_b(1.5\,\mbox{GeV}) = (4.63\pm 0.07)\,\mbox{GeV} \,, \qquad
   \mu_\pi^2(1.5\,\mbox{GeV}) = (0.15\pm 0.06)\,\mbox{GeV}^2 \,,
\end{equation}
where the errors reflect the uncertainties in the kinetic-scheme parameters 
only, and the correlation is the same as in the kinetic scheme. Note that we 
solve the relation for $\mu_\pi^2(\mu_f)$ in (\ref{simplebeauty}) exactly, 
rather than inverting it to $O(\alpha_s^2)$. (In the latter case the result 
for $m_b$ would stay unchanged, whereas the value of $\mu_\pi^2$ would 
increase to $(0.18\pm 0.06)$\,GeV$^2$, which is inside the error range.) For 
reference, the corresponding results obtained at one-loop order are 
$m_b(1.5\,\mbox{GeV})=(4.57\pm 0.07)$\,GeV and 
$\mu_\pi^2(1.5\,\mbox{GeV})=(0.34\pm 0.06)$\,GeV$^2$. The significant impact 
of two-loop corrections in the case of $\mu_\pi^2$ is due to the terms 
proportional to $\mu_f^2$ in the relation for $\lambda_1$ in 
(\ref{simplebeauty}), which arise first at $O(\alpha_s^2)$.

The potential-subtracted mass at the scale $\mu_f=2$\,GeV has been determined 
from an analysis of the $b\bar b$ production cross section and the mass of the 
$\Upsilon(1S)$ resonance to be $m_b=(4.59\pm 0.08)$\,GeV \cite{Beneke:1999fe}. 
Using relations (\ref{simplebeauty}) and (\ref{PSmassdef}), as well as the 
value of $\mu_\pi^2$ given in (\ref{values}), we may use this information to 
obtain $m_b(1.5\,\mbox{GeV})=(4.66\pm 0.08)$\,GeV for the $b$-quark mass in 
the shape-function scheme, in excellent agreement with (\ref{values}). We may 
also use the value $m_b=(4.68\pm 0.03)$\,GeV obtained in the $\Upsilon(1S)$ 
scheme from a fit to moments of $B\to X_c\,l\,\nu$ and $B\to X_s\gamma$ decay 
distributions \cite{Bauer:2004ve}, even though we do not trust the very small
error assigned in this analysis. Taking $\mu=1.5$\,GeV for the scale in 
(\ref{1Smassdef}), we find $m_b(1.5\,\mbox{GeV})=(4.61\pm 0.03)$\,GeV for the 
$b$-quark mass in the shape-function scheme, which is again consistent with 
(\ref{values}).

\begin{figure}
\begin{center}
\epsfig{file=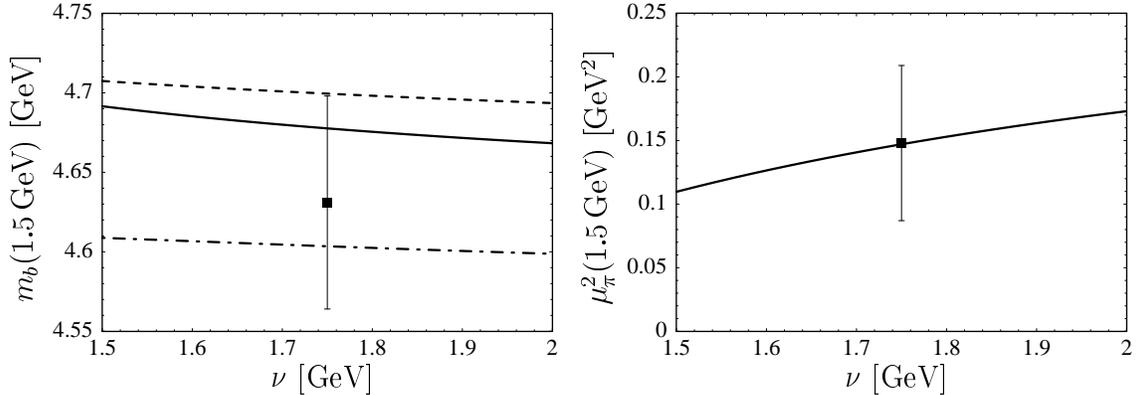,width=15.2cm}
\end{center}
\vspace{-0.2cm}
\centerline{\parbox{15cm}{\caption{\label{fig:scaledep}
Dependence of the heavy-quark parameters in the shape-function scheme on the 
scale $\nu$ at which the coupling constant is evaluated. The lines refer to 
the translation of results obtained in the kinetic scheme (solid), the 
potential-subtraction scheme (dashed), and the $\Upsilon(1S)$ scheme 
(dashed-dotted). Central values for the heavy-quark parameters in these 
schemes are used as input. The points show the values in (\ref{values}) for 
comparison.}}}
\end{figure}

The numbers presented above are affected by additional theoretical 
uncertainties inherent in the translation from one scheme to another. A way of 
assessing these uncertainties is to evaluate the coupling constant 
$\alpha_s(\nu)$ in the various relations at a common scale $\nu$ different 
from the individual ``natural'' scales. This can be done using
\begin{equation}
   \alpha_s(\mu) = \alpha_s(\nu) \left[ 1 + \frac{\beta_0}{2}\,
   \frac{\alpha_s(\nu)}{\pi}\,\ln\frac{\nu}{\mu} + \dots \right] .
\end{equation} 
While the results are formally independent of the choice of $\nu$, the 
variations due to truncation effects may serve as an indicator of the impact 
of higher-order effects. Varying $\nu$ between 1.5 and 2\,GeV so as to cover 
the range of ``natural'' scales in the various schemes yields the results 
shown in Figure~\ref{fig:scaledep}. The perturbative uncertainties indicated 
by this analysis are somewhat smaller than the parameter uncertainties shown in 
(\ref{values}). Combining the two sources of errors, we quote our final 
results as
\begin{equation}
   m_b(1.5\,\mbox{GeV}) = (4.63\pm 0.08)\,\mbox{GeV} \,, \qquad
   \mu_\pi^2(1.5\,\mbox{GeV}) = (0.15\pm 0.07)\,\mbox{GeV}^2 \,.
\end{equation}
Ultimately, the shape-function parameters should be determined directly from 
experimental data, without a detour through another subtraction scheme. 
Moments of the photon energy spectrum in $B\to X_s\gamma$ decay are ideally 
suited for this purpose and will allow an extraction of $m_b$ with far better 
precision \cite{inprep}.

Given values of the heavy-quark parameters $m_b$ and $\mu_\pi^2$ defined in 
the shape-function scheme at some reference scale $\mu_f$, it is 
straightforward to solve the relations (\ref{simplebeauty}) to find the 
corresponding running parameters at a different scale. There is nothing 
unphysical about this scale dependence, which indeed is an essential feature 
of our approach. For instance, we obtain 
$m_b(1.1\,\mbox{GeV})=(4.64\pm 0.08)$\,GeV and 
$\mu_\pi^2(1.1\,\mbox{GeV})=(0.18\pm 0.07)$\,GeV$^2$ at a lower scale, which 
is appropriate for applications the $B\to X_s\gamma$ photon spectrum 
\cite{Neubert:2004dd,inprep}, and $m_b(2\,\mbox{GeV})=(4.60\pm 0.08)$\,GeV and 
$\mu_\pi^2(2\,\mbox{GeV})=(0.11\pm 0.07)$\,GeV$^2$ at a higher scale, which 
could be used to analyse moments of $B\to X_c\,l\,\nu$ decay distributions. 
More important evolution effects are encountered when the 
scales $\mu_f$ and $\mu$ are taken differently, in which case the relations 
(\ref{beauty}) must be used to compute the effects of scale changes. For 
example, with $\mu_f=1.1$\,GeV and $\mu=1.5$\,GeV we get
$m_b(1.1\,\mbox{GeV},1.5\,\mbox{GeV})=(4.58\pm 0.08)$\,GeV and 
$\mu_\pi^2(1.1\,\mbox{GeV},1.5\,\mbox{GeV})=(0.36\pm 0.06)$\,GeV$^2$.

\paragraph{\em Conclusions.}
In summary, we have derived the two-loop relations for the $b$-quark mass 
$m_b$ and kinetic-energy parameter $\mu_\pi^2$ in the shape-function scheme 
introduced in \cite{Bosch:2004th}. These relations are important for 
performing high-precision studies of $B$-decay observables. We have determined 
values for $m_b(\mu_f)$ and $\mu_\pi^2(\mu_f)$ at a reference scale 
$\mu_f=1.5$\,GeV by using as input the corresponding values of heavy-quark 
parameters in other subtraction schemes. In the future, the shape-function 
parameters should be determined directly from a fit to moments of the 
$B\to X_s\gamma$ photon spectrum.

\paragraph{\em Acknowledgments.}
We are indebted to Einan Gardi for many useful discussions, which have been 
instrumental in finding the (hopefully correct) expressions for the two-loop 
anomalous dimension in (\ref{littlegamma}). We are also grateful to Kolya 
Uraltsev for discussions on the kinetic scheme. This research was supported by 
the National Science Foundation under Grant PHY-0355005, and by the Department 
of Energy under Grant DE-FG02-90ER40542. 

\paragraph{\em Note added.}
After this work was submitted for publication, an Erratum to 
\cite{Korchemsky:1992xv} appeared, in which the authors confirm the result
(\ref{littlegamma}) with $\kappa=4/3$.


\begin{thebibliography}{29}
\bibitem{Einanprivate}
E.~Gardi reports that he has confirmed (\ref{littlegamma}) with $\kappa=4/3$ 
from an independent calculation of the anomalous dimension $\gamma$ (private 
communication).
\end{thebibliography}
\end{document}